\documentclass[letter,twocolumn]{jpsj2} 
\def\runauthor{Takafumi \textsc{Mizuno} \textit{et al}.}
\title{
Excitonic Instability in the Transition from the Black Phase to the Golden Phase of SmS under Pressure Investigated by Infrared Spectroscopy
}
\author{
Takafumi \textsc{Mizuno}$^1$, 
Takuya \textsc{Iizuka}$^1$, 
Shin-ichi \textsc{Kimura}$^{2,1,}$\thanks{E-mail: kimura@ims.ac.jp}, 
Kazuyuki \textsc{Matsubayashi}$^{3,}$\thanks{Present address: Institute for Solid State Physics, The University of Tokyo, Kashiwa 277-8581, Japan}, 
Keiichiro \textsc{Imura}$^3$, 
Hiroyuki S. \textsc{Suzuki}$^4$, 
and
Noriaki K. \textsc{Sato}$^3$
}
\inst{
$^{1}$School of Physical Sciences, The Graduate University for Advanced Studies (SOKENDAI), Okazaki 444-8585, Japan\\
$^{2}$UVSOR Facility, Institute for Molecular Science, Okazaki 444-8585, Japan\\
$^{3}$Department of Physics, Nagoya University, Nagoya 464-8602, Japan\\
$^{4}$National Institute for Materials Science, Tsukuba, Ibaraki 305-0047, Japan
}
\recdate
{
\today
}
\abst{
We report the pressure-dependent optical reflectivity spectra of a strongly correlated insulator, samarium monosulfide (SmS), in the far- and middle-infrared regions to investigate the origin of the pressure-induced phase transition from the black phase to the golden phase.
The energy gap becomes narrow with increasing pressure in the black phase.
A valence transition from Sm$^{2+}$ in the black phase to mainly Sm$^{3+}$ in the golden phase accompanied by spectral change from insulator to metal were observed at the transition pressure of 0.65~GPa.
The black-to-golden phase transition occurs when the energy gap size of black SmS becomes the same as the binding energy of the exciton at the indirect energy gap before the gap closes.
This result indicates that the valence transition originates from an excitonic instability.
}
\kword{SmS, excitonic instability, infrared spectroscopy, high pressure}
\begin{document}
\maketitle
%
In the translocation process from the local to itinerant character of carriers in strongly correlated electron systems, the physical properties drastically change due to the complex relation of the transport to magnetic properties.
Clarification of the origin of the anomalous properties from the electronic structure is currently an important subject in solid-state physics.
The target material dealt with in this paper, samarium monosulfide (SmS), has been actively studied from the initial stages of research on valence fluctuation because of its pressure-induced insulator-to-metal transition~\cite{Jaya1970}.
SmS is a semiconductor with an energy gap size of $\sim$~1000~K ($\sim$~90~meV) and its color is black (namely, the ``black phase'') at ambient pressure~\cite{Matsu2007-2}.
Above the critical pressure ($P_c$) of 0.65~GPa, the sample color changes to golden-yellow (the ``golden phase'') and the electrical resistivity then suddenly drops to one-tenth that in the black phase~\cite{Keller1979, Raymond2002}.
The volume also decreases by about 10~\% with the same NaCl-type crystal structure~\cite{Annese2006}.
As the volume changes, the electronic configuration of the Sm ion changes from divalence to mainly trivalence by the first-order transition~\cite{Annese2006}.
In the golden phase, recent specific heat measurements under pressure have revealed that a pseudo-gap with a gap size of about 100~K exists~\cite{Matsu2007-1}.
Golden SmS has therefore been attracting attention because the electronic structure seems to be related to that of a typical Kondo semiconductor, SmB$_6$~\cite{Kimura1994}.
The transition from the band insulator of black SmS to the mixed-valent semiconductor of golden SmS is a local-to-itinerant transition.
In this valence transition, it is important to determine whether the electronic structure continuously changes in TmTe~\cite{Jarrige2006} due to the closing of the energy gap between the Tm~$4f$ and $5d$ states or discontinuously changes in YbInCu$_4$~\cite{Mito2007} due to the valence instability to clarify the mechanism of the black-to-golden phase transition in SmS.
In addition, if the valence transition is discontinuous, clarifying the origin of the valence transition will provide important information for the local-to-itinerant transition in strongly correlated electron systems.

Although the differences in physical properties of golden SmS and black SmS have been clarified during the past more than 30 years, the origin of the black-to-golden phase transition has never been elucidated due to the lack of direct observation of the pressure-dependent electronic structure very near the Fermi level.
To investigate the electronic structure, optical measurements corresponding to the electronic structure have already been performed for many years~\cite{Kaldis1972, Batlogg1976, Gunt1983, Travaglini1984}.
Optical reflectivity measurements under pressure have also been performed~\cite{Kirk1972}.
The paper indicated that the plasma edge due to carriers suddenly appears at 3~eV above $P_c$, which suggests a rapid increase in carrier density.
The carrier density increase corresponds to the appearance of the Sm$^{3+}$ state in the golden phase replacing the Sm$^{2+}$ state in the black phase.
However, the translocation process has never been clarified because the measured spectral region was limited to the visible and near-infrared regions.

In this paper, we report on the optical reflectivity [$R(\omega)$] spectra of SmS under pressure not only in the middle-infrared but also in the far-infrared region, as well as our corresponding band structure calculations, to obtain information on the pressure-dependent electronic structure.
As a result of the present study, it was found that the black-to-golden phase transition occurs when the energy gap size of black SmS becomes the same as the binding energy of the exciton before the gap closes.
This suggests that the black-to-golden phase transition of SmS originates from an excitonic instability.

A high-quality single crystal of SmS (\#8 in Ref.~2) was grown by the vertical Bridgman method in a high-frequency induction furnace.
The energy gap was determined to be about $1080\pm50$~K ($93\pm4$~meV) by electrical resistivity measurement.
The far-infrared $R(\omega)$ spectra under pressure were obtained at the THz microspectroscopy end station of an infrared synchrotron radiation beamline BL6B of UVSOR-II, Institute for Molecular Science, Okazaki, Japan~\cite{Kimura2006}.
A diamond anvil cell was employed to apply high pressures to the samples.
A sample cleaved along the $(001)$ plane with a typical size of 0.4~$\times$~0.4~$\times$~0.05~mm$^3$ was set in the diamond anvil cell with a culet plane size of 1.0~mm in diameter, using Apiezon-N grease or KBr powder as a pressure medium, gold films on the sample and on the gasket as reflectivity reference, and ruby chips as pressure reference.
The pressure was calibrated by ruby fluorescence measurement.
The pressure distribution did not change below 1~GPa because no broadening of the ruby line was observed.
The pressure-dependent $R(\omega)$ spectra were measured only with increasing pressure.
We also measured the temperature dependence of the $R(\omega)$ of SmS in a wide energy range from 10~meV to 30~eV using synchrotron-based equipment at UVSOR-II as a reference.
The optical conductivity [$\sigma(\omega)$] spectra were derived from the Kramers-Kronig analysis of the $R(\omega)$ spectra after extrapolating using a constant below 10~meV and $R(\omega) \propto \omega^{-4}$ above 30~eV.

The calculated $\sigma(\omega)$ spectrum of the direct interband transition was obtained using the full-potential linearized augmented plane wave plus local orbital (LAPW+lo) method including spin-orbit coupling and the on-site Coulomb repulsion energy ($U$) implemented in the {\sc Wien2k} code~\cite{WIEN2k}.
The value of $U$ was set at 7~eV, which is a standard value for rare-earth compounds~\cite{Antonov2004}.
The calculated energy gap size at ambient pressure was about 1000~K.

\begin{figure}[t]
\begin{center}
\includegraphics[width=0.40\textwidth]{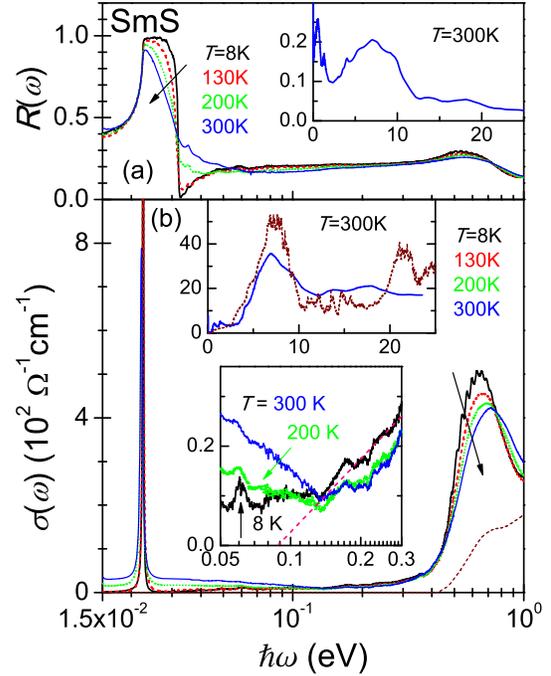}
\end{center}
\caption{
(Color online)
(a) Temperature dependence of reflectivity [$R(\omega)$] spectrum of SmS at ambient pressure.
The $R(\omega)$ spectrum in the wide energy region up to 25~eV is also plotted in the inset.
(b) Temperature dependence of optical conductivity [$\sigma(\omega)$] spectrum of SmS at ambient pressure derived from Kramers-Kronig analysis of the $R(\omega)$ in (a).
The dashed line is the calculated $\sigma(\omega)$ spectrum.
The upper inset shows the experimental $\sigma(\omega)$ (solid line) at 300~K and the calculated $\sigma(\omega)$ (dashed line) in the wide energy range.
The lower inset shows the enlargement near the indirect energy gap in the energy region of 0.05~--~0.3~eV at 8, 200 and 300~K.
The straight dashed line was fitted to the higher energy part, and the intersection with the abscissa indicates the indirect energy gap.
There is a peak at 60~meV indicated by an arrow, which originates from an exciton at the indirect energy gap.
}
\label{TdepR}
\end{figure}
In SmS at ambient pressure, since the Sm ion is divalent and the total magnetic moment $J=0$, the ground state is a nonmagnetic band insulator.
Corresponding to the fundamental physical properties, $R(\omega)$ has a large transverse optical (TO) phonon peak due to the stretching mode between Sm and S ions and the Drude structure due to carriers that does not appear at low temperatures, as shown in Fig.~\ref{TdepR}(a).
In the $R(\omega)$ spectra, the structure due to the interband transition in the energy range above 0.4~eV is similar to that reported in previous papers~\cite{Gunt1983}, but the spectra below that energy range are quite different because there is no Drude structure.
As reported in our previous paper, the carrier appears due to the shift from the stoichiometry~\cite{Matsu2007-2}.
Therefore, the sample used in this study is of higher quality from the point of view of the stoichiometry.
The $\sigma(\omega)$ spectra in Fig.~\ref{TdepR}(b) have gentle and steep slopes from about 0.09 and 0.4~eV to the higher energy side, respectively.
The former structure relates to the indirect transition from the top of the Sm~$4f$ state at the $\Gamma$-point to the bottom of the Sm~$5d$ band at the $X$-point and the latter to the direct transition of Sm~$4f \rightarrow 5d$ at the $X$-point according to the band structure calculated by using the LSDA+U method~\cite{Kimura2008}, which is basically consistent with the previously performed band calculation~\cite{Antonov2002} and with the angle-resolved photoemission spectrum~\cite{Ito2002}.
The experimental indirect energy gap size is about 90~meV, which is the intersection with abscissa of the straight line fitted with the higher energy side of $\sigma(\omega)$ spectrum at 8~K in the lower inset of Fig.~\ref{TdepR}(b).
The indirect energy gap size is consistent not only with the electrical resistivity data but also with the band calculation ($\Delta E\sim$1000~K~=~86~meV)~\cite{Kimura2008}.
The theoretical $\sigma(\omega)$ spectrum calculated using the band structure is also consistent with that obtained experimentally, as shown in the upper inset of Fig.~\ref{TdepR}(b).
Therefore, the optical spectra at ambient pressure can be fundamentally explained by the band structure calculation.
In the previous paper, the $\sigma(\omega)$ peak at 0.6~eV was assigned to originate from the $4f \rightarrow 5d$ interband transition~\cite{Batlogg1974}.
However, the experimental spectral shape of $\sigma(\omega)$ has strong temperature dependence and is conspicuously different from that obtained by calculation, in which the direct interband transition is assumed.
In addition, a vibronic-exciton structure originating from the coupling between the exciton and the longitudinal optical (LO) phonon~\cite{Sumi1975} (the period is 32~meV, which is the same as the LO phonon energy of the loss function peak obtained from the Kramers-Kronig analysis of the $R(\omega)$ spectrum in Fig.~\ref{TdepR}) appears in the 0.6-eV peak at 8~K~\cite{Kimura_unpublished}.
The 0.6-eV peak can thus be regarded to originate from the exciton due to the Sm~$4f \rightarrow 5d$ transition.

In the lower inset of Fig.~\ref{TdepR}(b), there is a peak at 60~meV, which is a lower energy than the indirect gap at $93\pm4$~meV.
Since the peak slightly shifts to the low-energy side with increasing temperature as shown in the figure and disappears with increasing defects (not shown), we suggest that the peak originates from the exciton of the indirect energy gap with a binding energy of about $33\pm4$~meV.

\begin{figure}[t]
\begin{center}
\includegraphics[width=0.40\textwidth]{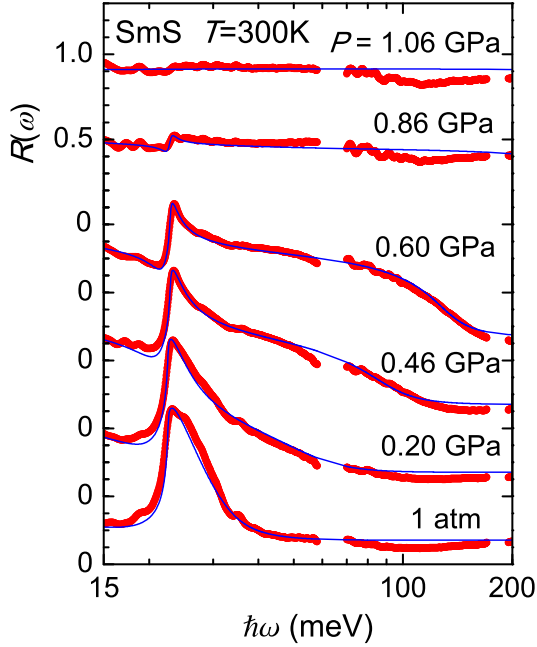}
\end{center}
\caption{
(Color online)
Pressure dependence of the reflectivity spectrum [$R(\omega)$] of black SmS (thick lines) in the energy region of $\hbar\omega$=~15~--~200~meV at 300~K.
The fitting curve of the combination of Drude and Lorentz functions (thin lines) are also plotted.
Successive curves are offset by 0.4 for clarity.
}
\label{PdepR_LowPE}
\end{figure}
\begin{figure}[t]
\begin{center}
\includegraphics[width=0.40\textwidth]{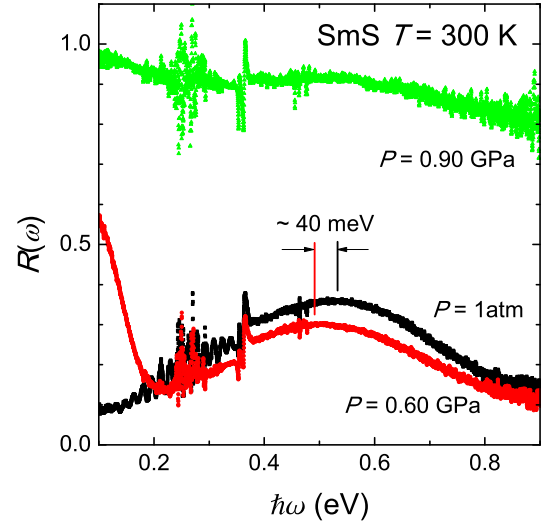}
\end{center}
\caption{
(Color online)
Pressure dependence of the reflectivity spectrum [$R(\omega)$] of black SmS (thick lines) in the energy region of $\hbar\omega$=~0.1~--~0.9~eV.
The vertical lines indicate the peaks at ambient pressure and 0.60~GPa.
The peak at $540\pm20$~meV at 1~atm slightly shifts to $500\pm20$~meV at 0.60~GPa.
}
\label{PdepR_HighPE}
\end{figure}
\begin{figure}[t]
\begin{center}
\includegraphics[width=0.40\textwidth]{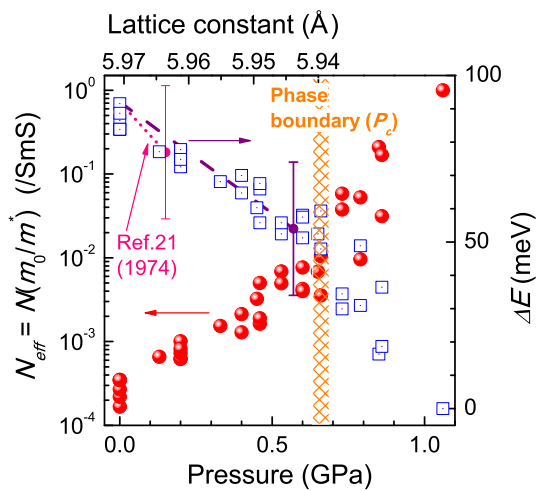}
\end{center}
\caption{
(Color online)
Pressure dependences of the effective carrier density ($N_{eff}$, solid circles) evaluated from the Drude and Lorentz fitting in Fig.~\ref{PdepR_LowPE} and energy gap size ($\Delta E$, open squares) evaluated from $N_{eff}$ and $\Delta E$ at 1~atm.
The lattice constant that is proportional to the pressure~\cite{Keller1979} is also denoted in the figure for reference.
The energy shift of the exciton peak in Fig.~\ref{PdepR_HighPE} (dashed line) and that of the previous data (dotted line)~\cite{Batlogg1974} normalized at the energy gap size at ambient pressure are plotted as well.
See the text for details.
}
\label{PdepParam}
\end{figure}

The pressure dependence of the $R(\omega)$ of SmS at 300~K in the far-infrared region is shown in Fig.~\ref{PdepR_LowPE}.
At ambient pressure, the overall $R(\omega)$ is low except for the phonon structure at 24~meV.
The spectrum can be fitted by the same parameters of the Drude and Lorentz functions~\cite{DG} as were used in the fitting of the temperature-dependent $R(\omega)$ in Fig.~\ref{TdepR}(a) with the refractive index of diamond ($n_0=2.37$)~\cite{Palik1985}.
When pressure is applied, the phonon peak does not shift but the $R(\omega)$ spectral weight increases.
This indicates that the Drude weight as well as the carrier density increases with pressure.
The fitting curves with Drude and Lorentz functions are also plotted in Fig.~\ref{PdepR_LowPE} and the effective carrier density [$N_{eff}=N(m_0/m^*)$, where $N$ is the carrier density, $m_0$ the rest mass of an electron, and $m^*$ the effective mass of the carrier] obtained from the fitting results is plotted by solid circles in Fig.~\ref{PdepParam}.
The fitting function is as follows:
\[
\hat{\varepsilon}(\omega)=\varepsilon_{\infty}+\frac{4\pi e^2}{m_0}\left[\frac{-N_{eff}}{\omega^2+i\omega\Gamma}
+\frac{N_p}{(\omega_p^2-\omega^2)-i\omega\Gamma_p}\right],
\]
\[
R(\omega) = \left|\frac{n_0-\hat{\varepsilon}(\omega)^{1/2}}{n_0+\hat{\varepsilon}(\omega)^{1/2}}\right|^{2},
\]
where, $\hat{\varepsilon}(\omega)$ is a complex dielectric function; 
$\varepsilon_{\infty}$ is the sum of $\varepsilon_{1}$ above the measured energy region;
$N_{eff}$ and $\Gamma$ the effective number and the scattering rate of the carriers, respectively;
and $N_p$, $\omega_p$ and $\Gamma_p$ are the intensity, resonance frequency, and the scattering rate of the TO-phonon, respectively.
In the fitting, although the parameters of the phonon peak [$\hbar\omega_p$~=~22.5--22.9~meV, $\hbar\Gamma_p$~=~0.18--0.21~meV, $N_p$~=~(0.98--2)$\cdot10^{-4}$~/SmS from 1~atm to 0.6~GPa], of the $\hbar\Gamma$ of the carriers (44.2~meV) and of the $\varepsilon_{\infty}$ (7.5) do not significantly change, the $N_{eff}$ of the Drude part changes.
The measurement was performed four times and all fitting data are plotted in Fig.~\ref{PdepParam}.
As shown in the figure, $N_{eff}$ exponentially increases with increasing pressure up to $P_c$.
Since the band dispersion near the $E_{\rm F}$ does not change so much with decreasing lattice constant by the pressure-dependent band structure calculation of SmS, the $m^*$ is considered to be a constant with increasing pressure.~\cite{Kimura_unpublished}
Then $N$ as well as $N_{eff}$ is proportional to $\exp(-\Delta E/k_{\rm B}T)$, where $\Delta E$ is the energy gap and $k_{\rm B}$ the Boltzmann constant.
This indicates that $\log N_{eff} \propto -\Delta E/k_{\rm B}T$.
Figure~\ref{PdepParam} does, in fact, indicate that $\log N_{eff}$ is proportional to the pressure; {\it i.e.}, the energy gap size becomes narrow in proportion to the pressure.
It is noted that $N_{eff}$ diverges from the relation of $\log N_{eff} \propto -\Delta E/k_{\rm B}T$ above $P_c$.
However, $N_{eff}$ is not discontinuous at $P_c$ in spite of the first-order black-to-golden phase transition.
The reason is the phase separation of the black and golden phases, because a patchy pattern can be visually observed (not shown).
In addition, at 0.90~GPa in Fig.~\ref{PdepR_HighPE}, the $R(\omega)$ spectrum appears to be a mixture of black and golden SmS.
This also indicates that the black-golden phase separation appears at this pressure.

From $\log N_{eff} \propto -\Delta E/k_{\rm B}T$ and the energy gap size of 93~$\pm$~4~meV at 1~atm, the pressure-dependent $\Delta E$ obtained from $N_{eff}$ can be evaluated as shown by open squares in Fig.~\ref{PdepParam}.
In the figure, the pressure-dependent exciton peak around 0.5~eV in Fig.~\ref{PdepR_HighPE} is also plotted.
In Fig.~\ref{PdepR_HighPE}, the peak observed at $540\pm20$~meV at ambient pressure slightly shifts to $500\pm20$~meV at 0.60~GPa.
The value of the peak shift ($-67\pm20$~meV/GPa) is almost consistent with that reported in the previous paper ($-100\pm20$~meV/GPa) measured below 0.15~GPa~\cite{Batlogg1974}, which is also plotted in Fig.~\ref{PdepParam}.
The exciton peak energies of these data are normalized at the energy gap size at ambient pressure, {\it i.e.}, the exciton peak energy obtained from Fig.~\ref{PdepR_HighPE} is shifted by -40~$\pm$~20~meV at 0.6~GPa in Fig.~\ref{PdepParam}.
The velocity of the peak shift at 0.5~eV is consistent with $\Delta E$ as shown in the figure.
This means that the energy gap becomes narrow without deformation of the electronic structure.
Note that the peak energy at 1~atm in Fig.~\ref{PdepR_HighPE} is not consistent with that at 300~K in Fig.~\ref{TdepR}(a) because the $R(\omega)$ spectrum in Fig.~\ref{PdepR_HighPE} was taken in the diamond anvil cell in which $n_0$ is 2.37 that is different from that in Fig.~\ref{TdepR}(a).

As shown in Fig.~\ref{PdepParam}, the energy gap size at $P_c$ is $45\pm20$~meV.
This means that the energy gap in the black phase does not close at $P_c$.
The energy gap size is close to the binding energy ($33\pm4$~meV) of the exciton at $\hbar\omega\sim$~60~meV in Fig.~\ref{TdepR}(b).
In other words, the insulator-to-metal transition occurs when the energy gap size becomes similar to the binding energy of the exciton before the gap closes.
This result is similar to the scenario of an excitonic insulator~\cite{Zhito1999, Jacques1965}.
This theory predicts that electrons and holes created from an overlap behave as a gas of oppositely charged particles, and coulombic attraction between electrons and holes can lead to spontaneous condensation of bound exciton pairs~\cite{Zhito1999}.
Then, in the case of SmS, the electrons in the Sm$^{2+}~4f$ valence band mix with the $5d$ conduction band.
As a result, a part of the Sm$^{2+}$ state changes to the Sm$^{3+}$ state.
Because of the smaller ionic radius of the Sm$^{3+}$ ion, the lattice constant becomes small after the change from Sm$^{2+}$ to Sm$^{3+}$ takes place.
The smaller lattice constant induces the electrons in the Sm$^{2+}~4f$ state to intermittently move to the $5d$ state.
Finally, almost all of the Sm$^{2+}~4f$ electrons change to Sm$^{3+}$.
This is one possible scenario for the black-to-golden phase transition in SmS.

In the theoretical prediction, a new state called an excitonic insulator is created above $P_c$~\cite{Zhito1999}.
The golden SmS also has a pseudo-gap with a size of about 100~K above $P_c$ as revealed by recent specific heat measurements~\cite{Matsu2007-1}.
One possibility is that the pseudo-gap state might be due to an excitonic insulator.
A similar energy gap that is believed to originate from an excitonic insulator also appears in La-doped SmS~\cite{Wachter1995}.
To clarify the origin of the pseudo-gap state in golden SmS, the far-infrared $R(\omega)$ of golden SmS must be investigated under pressure and at low temperatures.

In conclusion, far- and mid-infrared reflectivity measurements of SmS were performed under pressure.
At ambient pressure, the reflectivity spectrum indicates a perfect insulator in which Sm$^{2+}$ is in the black phase with indirect and direct gaps at about 90~meV and 0.4~eV, respectively, and with an exciton peak at 60~meV.
With increasing pressure, the energy gap size becomes narrow in proportion to the pressure.
At the transition pressure of 0.65~GPa, the insulating electronic structure of black SmS drastically changes to the metallic electronic structure of golden SmS.
However, the energy gap does not close but the size becomes similar to the binding energy of the exciton at the transition pressure.
The origin of the black-to-golden phase transition is therefore concluded to be due to an excitonic instability.

We would like to thank Prof. Kayanuma for his suggestion of the exciton of SmS.
This work was a joint studies program of the Institute for Molecular Science (2006) and was supported in part by a Grant-in-Aid of Scientific Research (B) (No.~18340110) from MEXT of Japan.


\end{document}